\documentclass[%
superscriptaddress,
 amsmath,amssymb,
 aps,
 notitlepage,
 twocolumn,
prb,
]{revtex4-2}

\usepackage{graphicx}% Include figure files
\usepackage[version=4]{mhchem}
\usepackage[colorlinks=True,linkcolor=red,citecolor=blue,urlcolor=blue]{hyperref}
\usepackage[utf8]{inputenc}
\newcommand{\bacu}{\ce{BaCuTe2O6}}

\graphicspath{{figs/}}

\begin{document}

\title{Magnetic excitation spectrum and Hamiltonian of the quantum spin chain \ce{BaCuTe2O6}}

%\author{authors}
%\affiliation{and their affiliations}
\author{A. Samartzis}
%\altaffiliation{alexandros.samartzis@helmholtz-berlin.de}
\affiliation{\mbox{Helmholtz-Zentrum Berlin f\"{u}r Materialien und Energie GmbH, Hahn-Meitner Platz 1, D-14109 Berlin, Germany}}
\affiliation{\mbox{Institut f\"{u}r Festk\"{o}rperphysik, Technische Universit\"{a}t Berlin, Hardenbergstra{\ss}e 36, D-10623 Berlin, Germany}}

\author{S. Chillal}
\affiliation{\mbox{Helmholtz-Zentrum Berlin f\"{u}r Materialien und Energie GmbH, Hahn-Meitner Platz 1, D-14109 Berlin, Germany}}

\author{H. O. Jeschke}
\altaffiliation{jeschke@okayama-u.ac.jp}
\affiliation{\mbox{Research Institute for Interdisciplinary Science, Okayama University, Okayama 700-8530, Japan}}

\author{D. J. Voneshen}
\affiliation{\mbox{ISIS pulsed neutron and muon source, STFC Rutherford Appleton Laboratory, Oxfordshire OX11 0QX, UK}}
\affiliation{\mbox{Department of Physics, Royal Holloway University of London, Egham, TW20 0EX}}

\author{Z. Lu}
\affiliation{\mbox{School of Computing, Engineering and The Built Environment, Edinburgh Napier University, Edinburgh EH10 5DT, UK}}

\author{A. T. M. N. Islam}
\affiliation{\mbox{Helmholtz-Zentrum Berlin f\"{u}r Materialien und Energie GmbH, Hahn-Meitner Platz 1, D-14109 Berlin, Germany}}

\author{B. Lake}
\altaffiliation{bella.lake@helmholtz-berlin.de}
\affiliation{\mbox{Helmholtz-Zentrum Berlin f\"{u}r Materialien und Energie GmbH, Hahn-Meitner Platz 1, D-14109 Berlin, Germany}}
\affiliation{\mbox{Institut f\"{u}r Festk\"{o}rperphysik, Technische Universit\"{a}t Berlin, Hardenbergstra{\ss}e 36, D-10623 Berlin, Germany}}

\date{\today}

\begin{abstract}
The magnetic excitation spectrum and Hamiltonian of the quantum magnet \ce{BaCuTe2O6} is studied by inelastic neutron scattering (INS) and density functional theory (DFT). INS on powder and single crystal samples reveals overlapping spinon continuua – the spectrum of an antiferromagnetic spin-1/2 spin chain – due to equivalent chains running along the {\bf a}, {\bf b}, and {\bf c} directions.  Long-range magnetic order onsets below $T_{\rm N}=6.3$~K due to interchain interactions, and is accompanied by the emergence of sharp spin-wave excitations which replace the continuua at low energies. The spin-wave spectrum is highly complex and was successfully modelled achieving excellent agreement with the data. The extracted interactions reveal an intrachain interaction, $J_3=2.9$~meV, while the antiferromagnetic hyperkagome interaction $J_2$, is the sub-leading interaction responsible for coupling the chains together in a frustrated way. DFT calculations reveal a similar picture for \ce{BaCuTe2O6} of dominant $J_3$ and sub-leading $J_2$ antiferromagnetic interactions and also indicate a high sensitivity of the interactions to small changes of structure which could explain the very different Hamiltonians observed in the sister compounds SrCuTe$_2$O$_6$ and \ce{PbCuTe2O6}.
\end{abstract}
\maketitle

\section{\label{sec:introduction}Introduction}

\begin{figure}
	\includegraphics[width=1.0\linewidth]{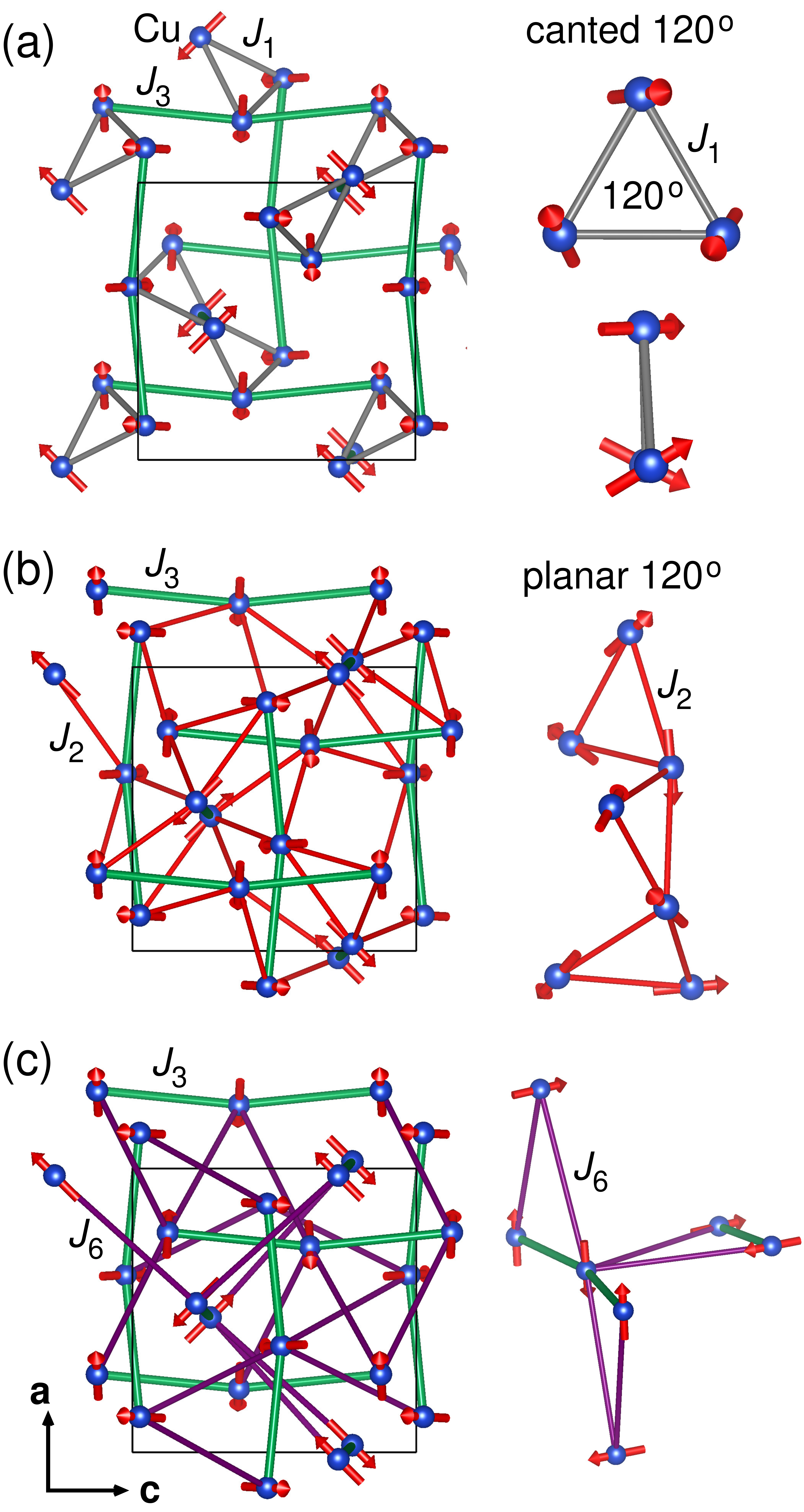}
	\caption{\label{fig:interactions}Visualization of the crystal structure of the ACuTe$_2$O$_6$ family showing only the  Cu$^{2+}$ ions which are represented by blue spheres. The crystal symmetry is cubic and the structure is viewed perpendicular to the {\bf b}-axis, where the black dashed lines indicate a unit cell. The colored lines indicate the different magnetic exchange interactions. The red arrows give the magnetic structure specific to \ce{BaCuTe2O6} \cite{Samartzis2021}. (a) Includes the first $J_1$ (grey line) and third $J_3$ (green line) nearest neighbor interactions. Note that $J_3$ produces equivalent chains parallel to the {\bf a}-, {\bf b}- and {\bf c}-axes due to the cubic crystal symmetry, while $J_1$ produces isolated triangles.  (b) Includes the second $J_2$ hyperkagome interaction given by the red line and the third $J_3$ interaction (green line). (c) Includes the $J_6$ interaction (purple line) which creates chains parallel to the body diagonals and the $J_3$ interaction (green line). On the right hand side of panels (a), (b) and (c), the triangles formed by $J_1$, $J_2$ and $J_3 \& J_6$ are presented respectively. This figure was produced using the VESTA software \cite{VESTA}. 
	}
\end{figure}

The compounds $A$CuTe$_2$O$_6$ ($A$ = Sr$^{2+}$, Pb$^{2+}$, Ba$^{2+}$) form a fascinating family of materials. Their properties include quantum magnetism, spin liquid behavior, ferroelectricity and magnetoelectricity. The magnetic properties arise from the Cu$^{2+}$ ions which have quantum spin-1/2 and are coupled by antiferromagnetic (AFM) interactions into chain and triangular structures. Meanwhile, the dielectric properties probably arise from the lone pair of electrons on the Te$^{4+}$ and Pb$^{2+}$ ions. The compounds form in the cubic space group $P4_{1}32$ (\#213) where all the Cu$^{2+}$ ions are equivalent and occupy the single $12d$ Wyckoff site. The Cu$^{2+}$ ions are coupled by a combination of one-dimensional (1D) and frustrated magnetic exchange interactions. As shown by Fig.~\ref{fig:interactions}, the first neighbor interaction $J_1$ couples them into isolated triangles, the second neighbor interaction $J_2$ couples them into a three-dimensional (3D) network of corner-sharing triangles known as the highly frustrated hyperkagome lattice and the third neighbor interaction, $J_3$ couples them into chains with equivalent chains parallel to the {\bf a}-, {\bf b}- and {\bf c}-axes due to the cubic symmetry. In addition, there are further neighbor interactions such as $J_4$ which also give rise to spin chains along the body diagonals and $J_6$ which couples together the parallel chains formed by the $J_3$ interaction.

Behavior consistent with a quantum spin liquid was found in \ce{PbCuTe2O6}. No indication of long-range magnetic order or spin freezing was found using a variety of experimental techniques even down to 20~mK \cite{Koteswararao2014,Khuntia2016,Chillal2020,Thurn2021,Hanna2021}. Instead, both nuclear magnetic resonance and muon spin resonance measurements suggested a non-spin-singlet ground state with persistent spin fluctuations \cite{Khuntia2016}, while inelastic neutron scattering showed broad diffuse excitations inconsistent with the spin-waves of a conventional magnet. Density functional theory calculations showed that the hyperkagome interaction $J_2$ is antiferromagnetic and strongest giving rise to strong magnetic frustration \cite{Koteswararao2014, Chillal2020} while the $J_1$ interaction was found to have similar strength \cite{Chillal2020}, leading to the novel hyper-hyperkagome lattice - a member of the distorted windmill lattice family \cite{Nakamura1997,HYKee2008}. Pseudo-Fermion functional renormalization group calculations based on these interactions confirm the absence of magnetic order and reproduce the observed magnetic structure factor. Further theoretical works using projective symmetry group analysis also suggest that \ce{PbCuTe2O6} could host quantum spin liquid states \cite{Jin2020,Chern2021}. More recently, interest has turned to a 1~K transition found in the single crystal samples which has been identified as a transition to ferroelectric order possibility related to the lone pairs on both the Pb$^{2+}$ and Te$^{4+}$ ions \cite{Thurn2021}. Interestingly, no long-range magnetic order was found so far to occur at this transition \cite{Thurn2021} and furthermore the transition is suppressed in powder samples because it is energetically unfavorable in small crystallites \cite{Hanna2021}. 
%%%%%%%%% \textcolor{red}{perhaps add references to the preprints of Fancelli, Eibisch and Hong.} %%%%%%%%%%%%%%%%%%%%%%%%%%%%%%%%%%%%%%%%

The behavior of SrCuTe$_2$O$_6$ is very different from \ce{PbCuTe2O6}. In this compound the unfrustrated $J_3$ interaction is dominant and antiferromagnetic giving rise to three equivalent spin chains parallel to the {\bf a}-, {\bf b}- and {\bf c}-axes. Evidence for this comes from broad peaks due to short-range order in the DC susceptibility and heat capacity \cite{Ahmed2015,Koteswararao2015, Koteswararao2016,Saeaun2020,Chillal2020_2}. Furthermore inelastic neutron scattering finds that the excitations form a spinon continuum characteristic of the spin-1/2 Heisenberg antiferromagnetic chain 
%which yields an intrachain interaction of strength $J_3=49$~K 
\cite{Chillal2021}. Physical properties measurements also reveal two transitions to long-range magnetic order and a complex phase diagram as a function of magnetic field and temperature \cite{ Ahmed2015,Koteswararao2015, Koteswararao2016, Saeaun2020, Chillal2020_2}. Furthermore measurement of the dielectric properties reveal magnetoelectric behaviour \cite{ Koteswararao2015, Koteswararao2016}. The long-range magnetic order in the ground state was studied by neutron diffraction and the magnetic structure is characterized by antiferromagnetic order along the $J_3$ chains and planar 120$^{\circ}$ order around the $J_1$ isolated triangles ($1\times\Gamma_1^1$ irreducible representation) \cite{Saeaun2020,Chillal2020_2}. Since long-range magnetic order is not possible in a one-dimensional magnet \cite{Mermin}, there must be additional magnetic interactions which couple the chains together. Several Density functional theory calculations were performed to calculate the interactions in SrCuTe$_2$O$_6$ which stated that the strongest interchain interaction is the hyperkagome interaction $J_2$ which was initially found to be antiferromagnetic (AFM) \cite{Ahmed2015, Koteswararao2015} but later found to be ferromagnetic (FM) \cite{Bag2021} along with a smaller AFM $J_6$ interaction \cite{ Ahmed2015,Bag2021}. The interchain interactions were also investigated by inelastic neutron scattering. By fitting the low energy magnetic spectrum at low temperatures to spin-wave theory, Chillal {\it et al} \cite{Chillal2021} showed that the FM $J_2$/AFM $J_6$ model is compatible with the data while an AFM $J_1$ also explains the spectrum.

The newest member of the ACuTe$_2$O$_6$ family is \ce{BaCuTe2O6}. As for SrCuTe$_2$O$_6$, the unfrustrated $J_3$ interaction is also dominant and antiferromagnetic in \ce{BaCuTe2O6}, giving rise to spin chains parallel to the {\bf a}-, {\bf b}- and {\bf c}-axes. The intrachain interaction was found to have strength $J_3=2.93-3.19$~meV from DC susceptibility and heat capacity measurements \cite{Samartzis2021,Bag2021}. This was confirmed by inelastic neutron scattering measurements which observed the characteristic spinon continuum expected for a spin-1/2 Heisenberg antiferromagnetic chain, allowing comparison to theory which yielded $J_3=2.90$~meV \cite{Samartzis2021}. Long-range antiferromagnetic order occurs at the much lower temperature of $T_{\rm N}=6.1-6.3$~K \cite{Samartzis2021,Bag2021} revealing the presence of weak interchain interactions. The magnetic structure was found to follow the $2\times\Gamma_2^1$ irreducible representation \cite{Samartzis2021}. It is illustrated by the red arrows along with the exchange interactions $J_1$, $J_2$, $J_3$ and $J_6$ in Fig.~\ref{fig:interactions}. As for SrCuTe$_2$O$_6$ the order is antiferromagnetic along the chains formed by $J_3$, however there is canted 120$^{\circ}$ order about the isolated $J_1$ triangles rather than the planar 120$^{\circ}$ order of SrCuTe$_2$O$_6$. Furthermore there is planar 120$^{\circ}$ order around the triangles of the hyperkagome lattice formed by $J_2$ (rather that canted 120$^{\circ}$ order). This suggests that an antiferromagnetic $J_2$ interaction is primarily responsible for coupling the $J_3$ chains and gives rise to the long-range magnetic order. Recent DFT calculations support this view, they predict that the chain interaction is $J_3=2.93$~meV, while $J_1=0.0$~meV and the interchain coupling is due to an AFM $J_2=0.43$~meV. They also predict an AFM $J_6=0.17$~meV while all other interactions are negligible \cite{Bag2021}.

\begin{figure}
	\includegraphics[width=0.8\linewidth]{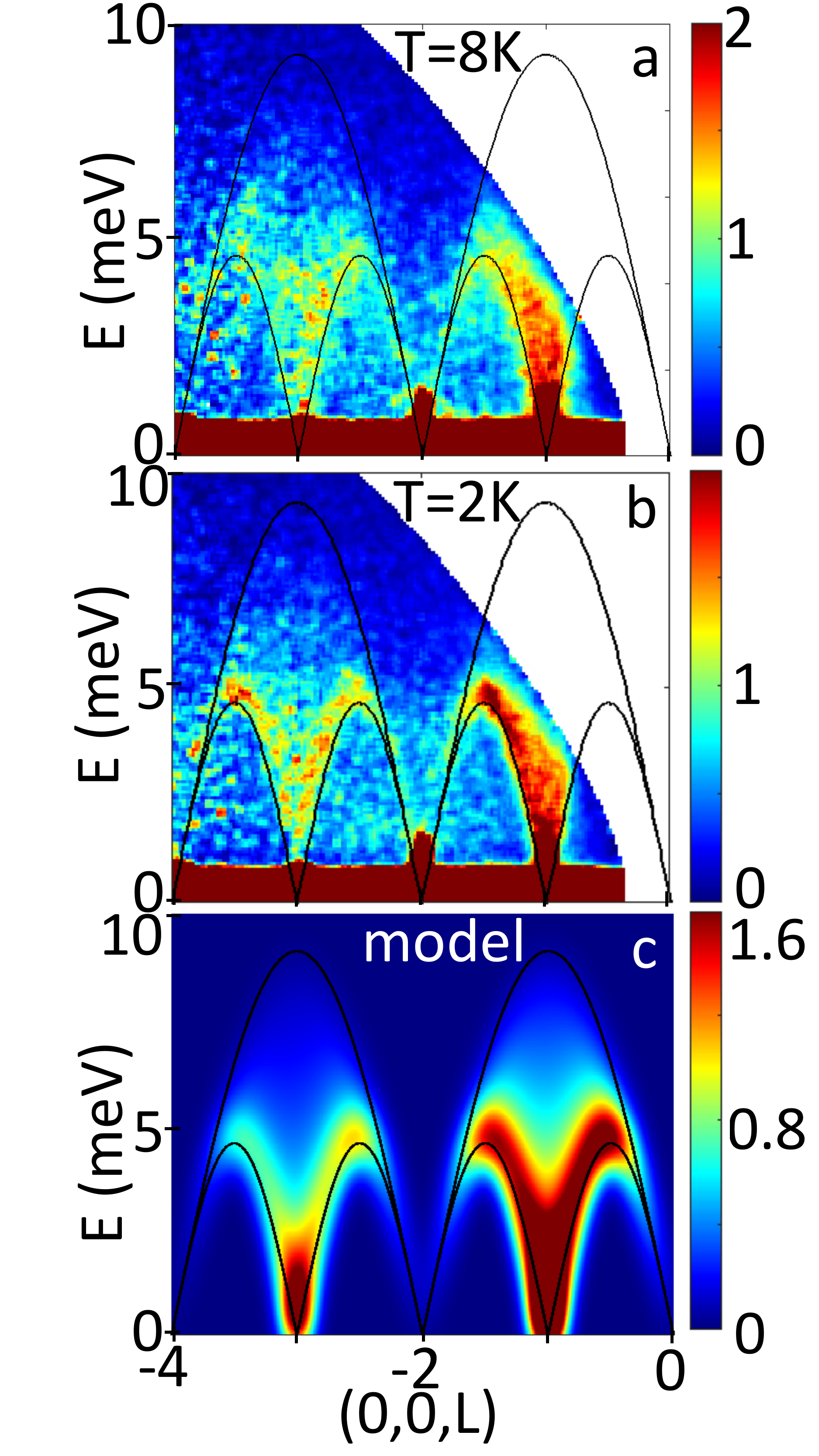}
	\caption{\label{fig:SpinonContinuum}High energy inelastic neutron scattering measured on the single crystal sample of \ce{BaCuTe2O6} compared to the theoretical spinon continuum. The data were collected on the LET spectrometer using an incident neutron energy of $E_{\text{i}}$=13.75~meV. The spectra are plotted as a function of energy and wavevector transfer along $[0,0,L]$ for (a) $T=8$~K ($>T_{\rm N}=6.3$~K) and (b) $T=2$~K ($<\ll T_{\rm N}=6.3$~K) respectively. They have been integrated by $\pm0.2$~r.l.u.\ over both the $[H,H,0]$ and $[K,-K,0]$ directions and no background has been subtracted. (c) Dynamical structure factor calculated via the algebraic Bethe-Ansatz for $J_3=2.90(6)$~meV \cite{Caux2005}. The magnetic form factor of the Cu$^{2+}$ ion was also included in the calculations and the data have been convolved with a Gaussian of width 0.78~meV to reproduce the effect of the energy resolution. The black lines plotted over the data and simulation represent the upper and lower boundaries of the spinon continuum.}
\end{figure}

\begin{figure}
	\includegraphics[width=0.82\linewidth]{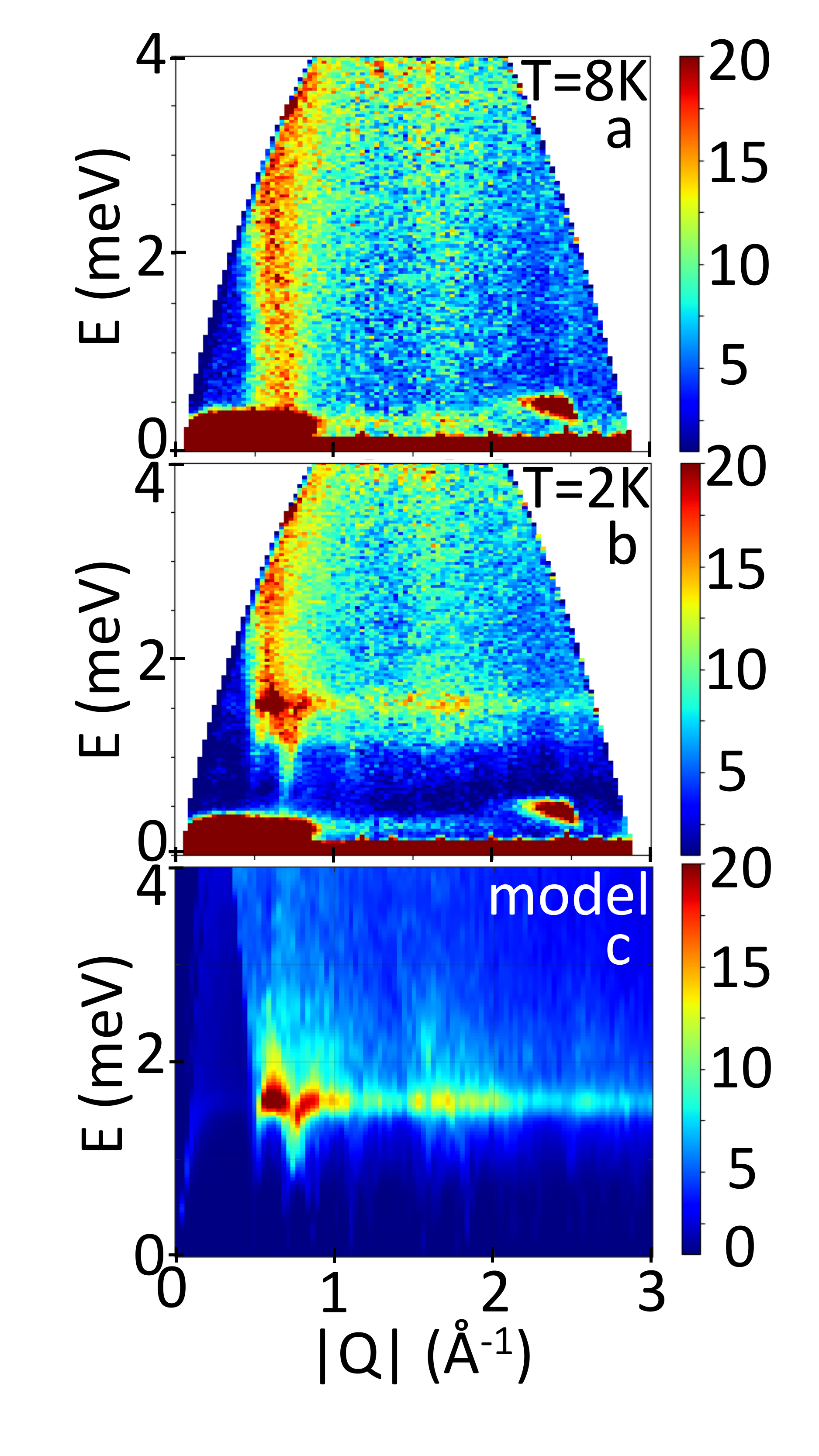}
	\caption{\label{fig:LETPowderEQ}Low energy powder inelastic neutron scattering spectrum displayed as a function of energy and wavevector transfer. The data were measured using the LET spectrometer at (a) $T=8$~K ($>T_{\rm N}=6.3$~K) and (b) $T=2$~K ($\ll T_{\rm N}=6.3$~K) with an incident energy of $E_{\text{i}}=4.88$~meV and energy resolution $\Delta E_{\text{i}}=0.144$~meV. No background has been subtracted from the data. Note that the data below 0.5~meV are unreliable showing several spurious features. (c) shows the powder spectrum simulated by spin-wave theory for $J_3'=\pi J_3/2$ where $J_3=2.90$~meV, $J_2=0.30$~meV, $J_1=0.05$~meV and $J_6=0.035$~meV. The magnetic form factor of the Cu$^{2+}$ ion was also included. The simulations were convolved over energy by a Gaussian function with a full-width-at-half-maximum (FWHM) of 0.144~meV to reproduce the broadening due to the instrumental resolution.}
\end{figure}

\begin{figure*}
	\includegraphics[width=0.9\linewidth]{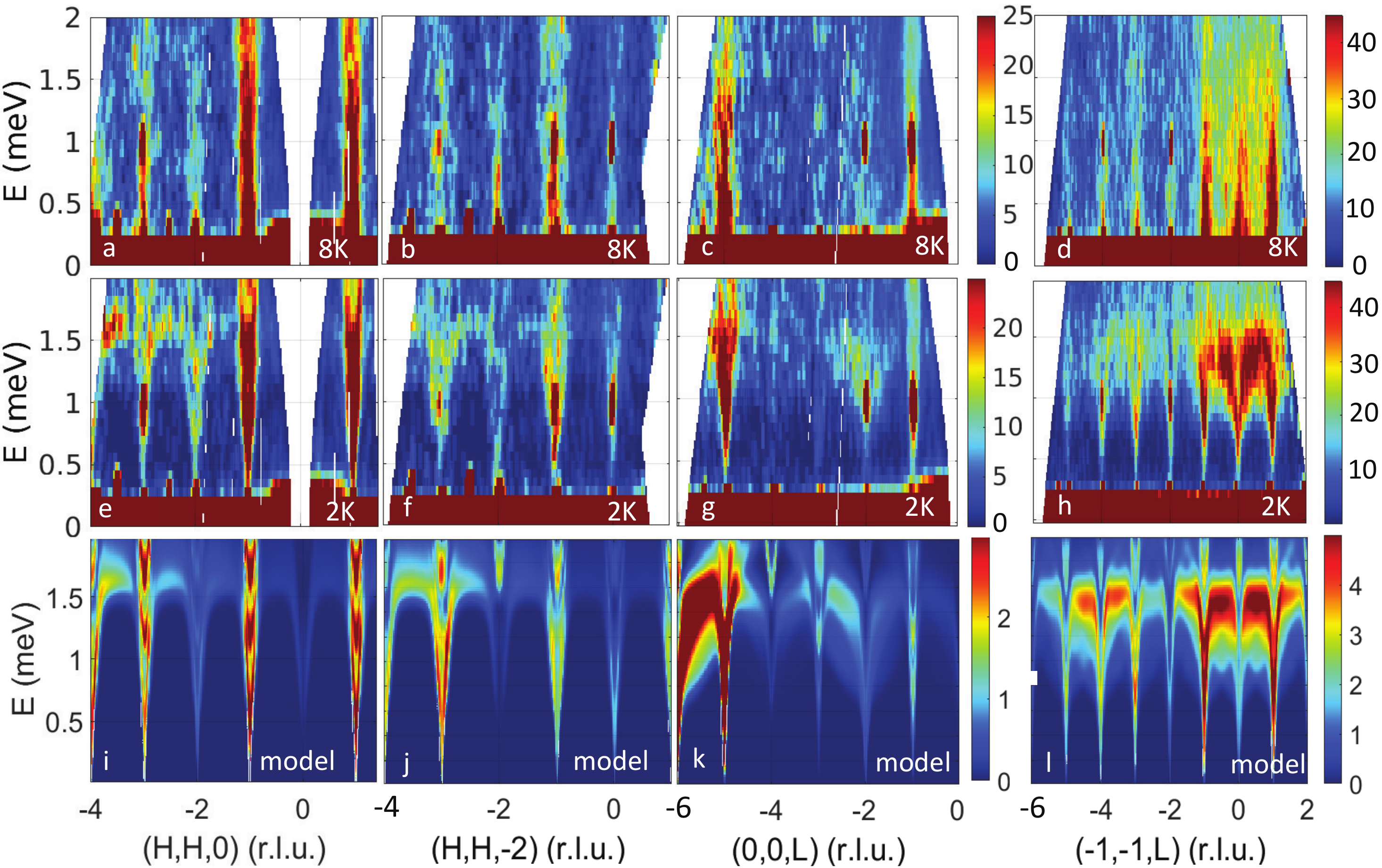}
	\caption{\label{fig:LETscEQCollection}Low energy inelastic neutron scattering data measured on the single crystal sample and plotted as a function of energy and selected wavevector transfers. The data were collected on the LET spectrometer using an incident neutron energy of $E_{\text{i}}=4.81$~meV. They are integrated by $\pm 0.1$~r.l.u.\ over both the vertical, out-of-plane scattering direction $[K,-K,0]$ and the horizontal direction perpendicular to the slice. (a), (b), (c) and (d) show the spectra along $[H,H,0]$, $[H,H,-2]$, $[0,0,L]$ and $[-1,-1,L]$ at $T=8$~K, while (e)-(h) show the spectra below $T_{\rm N}$ at $T=2$~K for the same directions. (i)-(l) shows the corresponding dynamical structure factor calculated by linear spin wave theory for interaction values of $J_3'=\pi J_3/2$ where $J_3=2.90$~meV, $J_2=0.30$~meV, $J_1=0.05$~meV and $J_6=0.035$~meV which is integrated over the same wavevector ranges as the data. The simulations were convolved over energy by a Gaussian function with FWHM of 0.175~meV to reproduce the broadening due to the instrumental resolution. The magnetic form factor of the Cu$^{2+}$ ion was also included. 
	%Details for the Spin-W: step of integrations oop-step=41, inp-step=41, integration oop-int=$\pm0.1$rlu, inp-int$\pm$0.1rlu.
	}
\end{figure*}

In this paper we determine the magnetic exchange interactions of \ce{BaCuTe2O6} using two independent methods. We extract the interactions from the magnetic excitation spectrum observed at low temperatures and energies using inelastic neutron scattering (INS). We also perform density functional theory (DFT) calculations to estimate the interactions based on the crystal structure. The results are compared to those of SrCuTe$_2$O$_6$ and \ce{PbCuTe2O6}.

\begin{figure*}
    \includegraphics[width=0.9\linewidth]{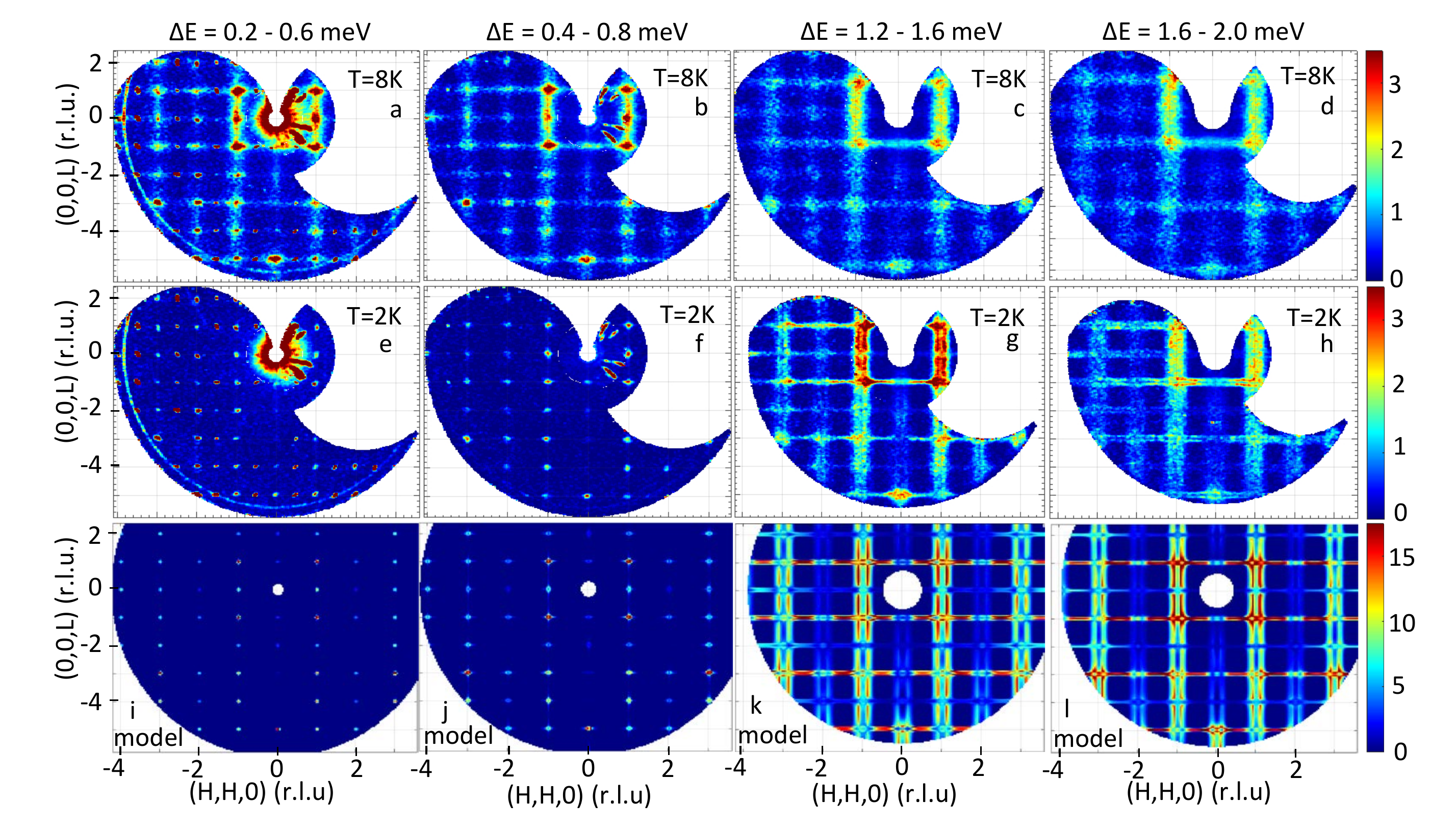}
	\caption{\label{fig:LETscQQCollection}Low energy inelastic neutron scattering data measured on the single crystal sample plotted as a function of wavevector transfer in the $[H,H,L]$-plane for various energy integration ranges which are given in the titles. The data were measured using the LET spectrometer at (a)-(d) $T=8$~K and (e)-(h) $T=2$~K using an incident energy of $E_{\text{i}}=4.81$~meV. The data were integrated over wavevector in the vertical out-of-plane direction $[K,-K,0]$ by $\pm 0.2$~r.l.u. No background subtraction has been performed and only a weak smoothing is implemented. The strong non-symmetric features at low energies and wavevector are spurious background from the instrument. (i)-(l) show the corresponding single crystal spectrum simulated by spin-wave theory for $J_3'=\pi J_3/2$ where $J_3=2.90$~meV, $J_2=0.30$~meV, $J_1=0.05$~meV and $J_6=0.035$~meV and integrated over the same wavevector and energy ranges as the data. The simulations were convolved over energy by a Gaussian function with FWHM of 0.175~meV to reproduce the broadening due to the effects of instrumental resolution. The magnetic form factor of the Cu$^{2+}$ ion was also included.
	%Details for the Spin-W: step of integrations oop-step=41, integration oop-int=$\pm0.2$rlu,
	}
\end{figure*}

\section{\label{sec:spectrum}Magnetic Excitation Spectrum}

\subsection{\label{sec:methods}Experimental Details}

Powder and single crystal samples of \ce{BaCuTe2O6} were synthesized and characterized at the Core Laboratory Quantum Materials, Helmholtz-Zentrum Berlin f\"{u}r Materialien und Energie, Germany. Further details can be found in Ref.\ \cite{Samartzis2021}. 

The spin dynamics of \ce{BaCuTe2O6} were studied by inelastic neutron scattering (INS). Data for both powder and single crystal samples were collected on the LET, cold neutron multi-chopper Time-of-Flight spectrometer operating at Target Station 2 of the ISIS facility, Rutherford Appleton Laboratory, UK \cite{LET,doiLET}. The samples were cooled using an orange cryostat and data were measured at $T$=2~K and 8~K. The spectrum of the powder sample (mass 10.1(4)~g) was collected simultaneously at five independent incident energies $E_\text{i}$= 14.1, 4.88, 2.45, 1.47 and 0.978~meV with corresponding elastic energy resolutions, $\Delta E_\text{i}$= 0.697, 0.144, 0.053, 0.025 and 0.014~meV for a total current of 240~$\mu$A and 310~$\mu$A at low and high temperatures, respectively. The single crystal sample consisted of three co-aligned crystals (total mass 4.80(1)~g) oriented with the $[1,-1,0]$ crystallographic direction vertical.  The crystals were wrapped in Al foil and attached to an Al sample holder using Al wire. This sample was measured at $T=2$~K and $T=8$~K with incident energies, $E_{\text{i}}$= 13.75, 7.6, 4.81, 3.32 and 2.43~meV and corresponding elastic energy resolutions, $\Delta E_{\text{i}}$= 0.782, 0.333, 0.175, 0.105 and 0.068~meV. During the measurements the crystals were rotated about their vertical axis over an angular range of 159$^{\circ}$ in steps of  0.5$^\circ$  or 1$^\circ$ for a current of 8~$\mu$A per step. Data processing and visualization were carried out using the Horace software \cite{Horace}.

Further INS measurements were performed to provide information about the low-energy magnetic excitations below the transition temperature. The co-aligned system of \ce{BaCuTe2O6} single crystals was measured again on the single detector, Triple-Axis-Spectrometer FLEXX, at Helmholtz-Zentrum Berlin f\"{u}r Materialien und Energie, Germany \cite{Flexx}. Data were collected at $T = 2$~K, using an orange cryostat. The wavevector of the scattered neutrons was fixed to $k_{\rm f} = 1.3$\, {\AA}$^{-1}$ which provided an elastic energy resolution of $\Delta E$=0.12~meV, while the monochromator was used in the horizontally and vertically focused mode to optimize intensity. Data were collected by scanning the energy with step size 0.05~meV, for fixed wavevectors along the $[1,1,L]$, $[0,0,L]$ and $[H,H,1]$ directions with values of $H$ and $L$ taken at intervals of 0.1~r.l.u.\ or smaller, measuring for $\approx 3.54$ min per step. In addition, $Q$-scans at constant-energy along the $[1,1,L]$ and $[H,H,1]$ directions were performed for a $Q$ step of 0.05~r.l.u.\ for several fixed energies with the same counting time. The visualization of the data was achieved using the software package Spec1D in Matlab.

The magnetic Hamiltonian of \ce{BaCuTe2O6}, was simulated using linear spin-wave theory within the Matlab package, SpinW \cite{Toth2015} and compared to the data at low energies and temperatures to obtain the exchange interactions.

\begin{figure*}
	\includegraphics[width=0.75\linewidth]{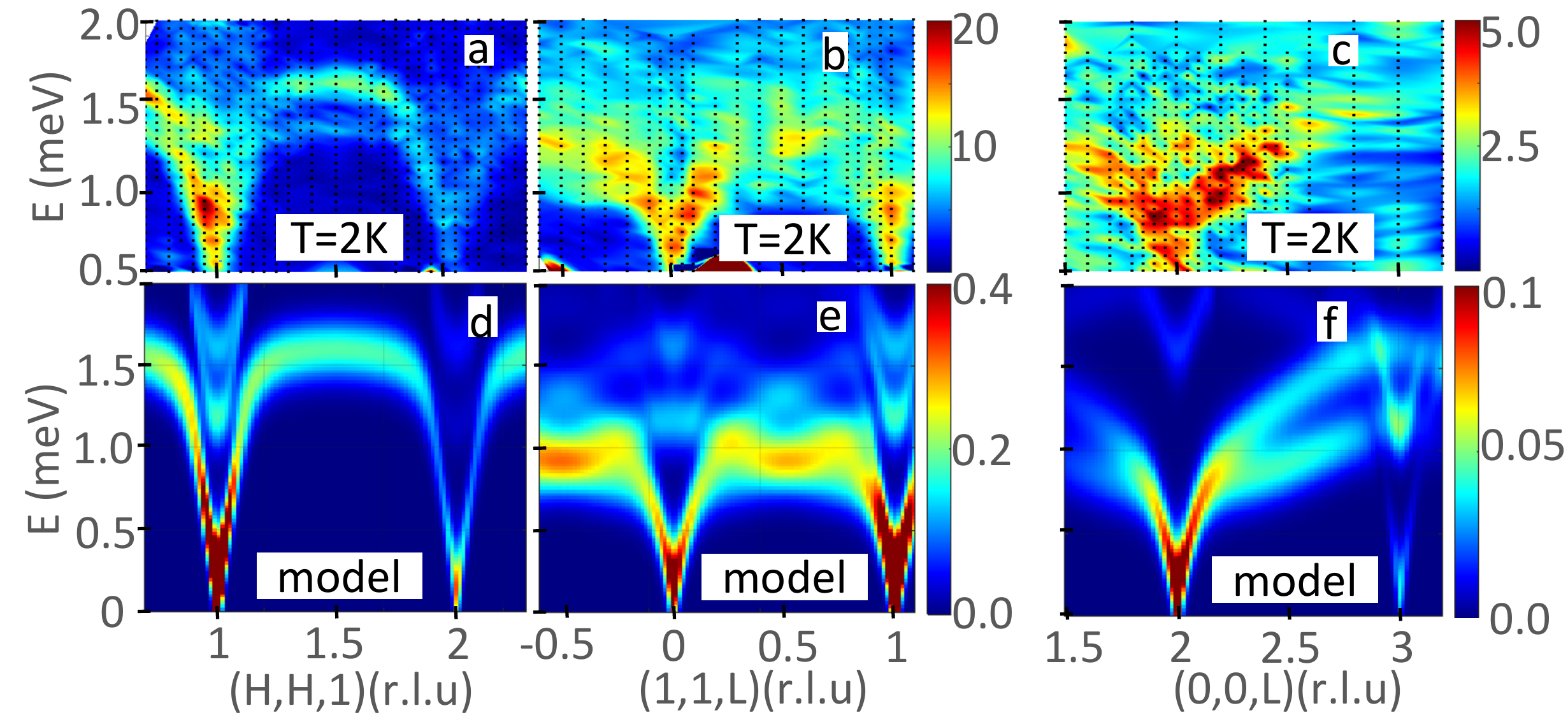}
	\caption{\label{fig:FLEXXscEQCollection}Single crystal inelastic neutron scattering spectra as a function of energy and wavevector transfer along the (a) $[H,H,1]$, (b) $[1,1,L]$ and (c) $[0,0,L]$-directions. The data were measured on the FLEXX spectrometer at $T=2$~K ($<T_{\rm N}=6.3$~K). No background has been subtracted. (d)-(f) show the corresponding single crystal spectra simulated by spin-wave theory for $J_3'=\pi J_3/2$ where $J_3=2.90$~meV, $J_2=0.30$~meV, $J_1=0.05$~meV and $J_6=0.035$~meV. 
 %The simulations were integrated over wavevector in the vertical out-of-plane direction $[K,-K,0]$ by $\pm0.1$~r.l.u and in the horizontal direction perpendicular each slice by $\pm0.02$~r.l.u to account for the out-of-plane and in-plane resolution respectively. 
 The wavevector resolution broadening was included by taking the Gaussian weighted average over the expected resolution widths in the directions out-of-plane ([$K$,$-K$,$0$]) and in-plane perpendicular to each slice by $0.1$~\AA$^{-1}$ and $0.02$~\AA$^{-1}$ respectively.
 The simulations were also convolved over energy by a Gaussian function of width 0.3~meV to mimic the broadening due to effects such as the instrumental energy resolution.
	%Details for the Spin-W: step of integrations oop-step=41, inp-step=15. The integration is weighted by a Gaussian rather than simply summed.
	}
\end{figure*}

\subsection{\label{sec:results}Results}

Figure~\ref{fig:SpinonContinuum} shows the full magnetic spectrum from the single crystal sample plotted as a function of energy up to 10~meV and wavevector along the direction of one set of chains due to the $J_3$ interaction. The data were measured with an incident energy of 13.75~meV on the LET spectrometer just above the N\'{e}el temperature at 8~K($>T_{\rm N}=6.3$~K) (Fig.~\ref{fig:SpinonContinuum}(a)) and well below it at 2~K($\ll T_{\rm N}$) (Fig.~\ref{fig:SpinonContinuum}(b)). Only subtle changes are observed with the decrease in temperature at most energy scales. As discussed in Ref.~\cite{Samartzis2021} the spectrum is typical of a spin-1/2, Heisenberg, antiferromagnetic chain showing a spinon continuum extending from a lower boundary given by the des Cloizeaux-Pearson expression $\left( \pi\times J_3 / 2 \right) \left| sin(\pi L)\right| $  to an upper boundary given by $\left( \pi\times J_3 \right) \left| sin(\pi L/2)\right|$~\cite{dCP}. This allowed the value of the intrachain interaction, $J_3=2.90$~meV, to be extracted \cite{Samartzis2021}. Figure~\ref{fig:SpinonContinuum}(c) shows the theoretical spinon continuum simulated for this value of $J_3$ for comparison with the data.

Differences in the spectrum above and below $T_{\rm N}$ are most pronounced at low energies. Figure~\ref{fig:LETPowderEQ} presents the low energy INS data collected on the LET spectrometer for the powder sample at 8~K (Fig.~\ref{fig:LETPowderEQ}(a)) and 2~K (Fig.~\ref{fig:LETPowderEQ}(b)) plotted as a function of energy and wavevector. At 8~K, there are streaks of scattering at the wavevectors 0.5 and 1.5~\AA$^{-1}$ which are due to the spinon continuum that comes down to low energies at (0,0,1) and (0,0,3) respectively. 
Note that these minima occur at odd-integer rather than half-odd-integer values of $L$ because each chain has two Cu$^{2+}$ ions per unit cell.
These are the same features observed in the high energy single crystal data shown in Fig.~\ref{fig:SpinonContinuum}. Below $T_{\rm N}$ the low energy spectrum is very different, a flat band appears at $\approx1.6$~meV and at lower energies new weak streaks appear at the wavevectors 0.5, 0.7 and 1.15~\AA$^{-1}$ which correspond to the magnetic Bragg peaks (0,0,1), (1,1,0) and (0,1,2)/(1,1,2) observed in neutron diffraction \cite{Samartzis2021}.

In order to learn more about the low energy spin dynamics of \ce{BaCuTe2O6}, further INS data were collected from the single crystal sample using lower incident energies on the LET spectrometer. Figure~\ref{fig:LETscEQCollection} presents the spectrum collected for an incident energy of $E_{\text{i}}$=4.81~meV plotted as a function of energy and wavevector for selected high symmetry directions. Above $T_{\rm N}$ at 8~K, strong magnetic signal is observed at odd integer values of $H$ and $L$ due to the spinon continuum, however at 2~K the signal below 2~meV is again significantly modified with a new dispersionless band at $\approx 1.6$~meV and the intensities at lower energies are dramatically altered. 

These changes can also be observed in the constant energy slices plotted in Fig.~\ref{fig:LETscQQCollection}. At 8~K (top panels of Fig.~\ref{fig:LETscQQCollection}) two sets of perpendicular streaks are observed. 
The streaks parallel to $[H,H,0]$, which are strongest for odd $H$, are due to the $J_3$ chains parallel to the {\bf c}-axis, these chains are effectively decoupled above $T_{\rm N}$ so that the dispersion perpendicular to these chains is negligible. Likewise the streaks parallel to $[0,0,L]$ are due to the chains along the {\bf a}- and {\bf b}-directions. Below $T_{\rm N}$ (middle panels of Fig.~\ref{fig:LETscQQCollection}) the excitations have clearly become dispersive. At the lowest energies, the streaks are replaced by dots suggesting spin-waves dispersing upwards from integer values of $H$ and $L$. Above $\approx 1.2$~meV the streaks are again visible, however they appear sharper than at 8~K and close observation shows them to be in fact double streaks, providing further evidence that these excitations are spin-waves.   

The results from the single crystal measurement on the FLEXX spectrometer are similar to the LET results. Figure~\ref{fig:FLEXXscEQCollection} shows the data at 2~K as a function of energy and wavevector along various directions. A corresponding measurement above $T_{\rm N}$ was not performed on this spectrometer.

In summary, while we can think of the chains due to the $J_3$ interaction as being essentially decoupled above $T_{\rm N}$ resulting in a spinon continuum in \ce{BaCuTe2O6}, at lower temperatures the interchain coupling gives rise to long-range magnetic order and spin-wave excitations at low energies. The higher energy spinon excitations are however almost unchanged by the transition. 
While the boundary between the spinon and spin-wave regimes is not sharp, the biggest temperature changes occur below $E \approx 1.6$~meV which is the top of the interchain spin-wave dispersion at low temperatures (see Fig.~\ref{fig:LETPowderEQ},
\ref{fig:LETscEQCollection} and \ref{fig:FLEXXscEQCollection}) identifying this as the spin-wave bandwidth.
Above $T_{\rm N}$ this dispersion is gone implying that the spin-waves (which have quantum spin number $S=1$) have 'deconfined' into pairs of spinons (which have quantum spin number $S=1/2$). 
In contrast, above $E \approx 2.0$~meV the excitations are broad and show almost no changes with temperature suggesting that this is the spinon regime. The intermediate energy regime ($1.6 < E < 2.0$~meV) is a crossover regime where the excitations can be described as broadened spin-waves or  partially confined spinons.

Such confinement of the spinon excitations into spin-waves at low energies and temperatures due to the interchain coupling  was observed previously and studied extensively in the spin-chain compound KCuF$_3$ \cite{Lake2000,Lake2005_2,Lake2005} and was also found in the sister compound SrCuTe$_2$O$_6$ \cite{Chillal2021}, which is isostructural to \ce{BaCuTe2O6}. 

\subsection{\label{sec:Analysis}Analysis}

Because we observe spin-waves excitations at low energies and temperatures in \ce{BaCuTe2O6}, the low energy INS data can be analyzed using linear spin-wave theory to extract the interchain exchange interactions. 
%The cross-over between the spin-wave and spinon energy regimes is not clearly defined, nevertheless the data suggests that sharp spin-waves exist below 1~meV and spin-wave theory is probably valid up to $\approx2$~meV while at higher energies the spinon model is more appropriate. 
The steepest spin-wave dispersion along the chain at low energies merges with the lower boundary of the spinon continuum at higher energies given by the des Cloizeaux-Pearson expression $(\pi\times J_3/2)\left|\sin(\pi H)\right|$ \cite{dCP}. Therefore, in order to model the dispersion along the chains using spin-wave theory, the renormalized intrachain exchange constant $J_3'=\pi\times J_3/2$ should be used. A consistent value of $2.84 < J_3 < 3.19$~meV has already been established in the literature from DC susceptibility, heat capacity, inelastic neutron scattering and DFT \cite{Samartzis2021,Bag2021}. Here we use $J_3=2.90$~meV derived by comparing our high energy INS data to the spectrum of the theoretical spinon continuum (Fig.~\ref{fig:SpinonContinuum}) \cite{Samartzis2021}, which gives us the value $J_3'=4.56$~meV.

\begin{figure}
	\includegraphics[width=0.9\linewidth]{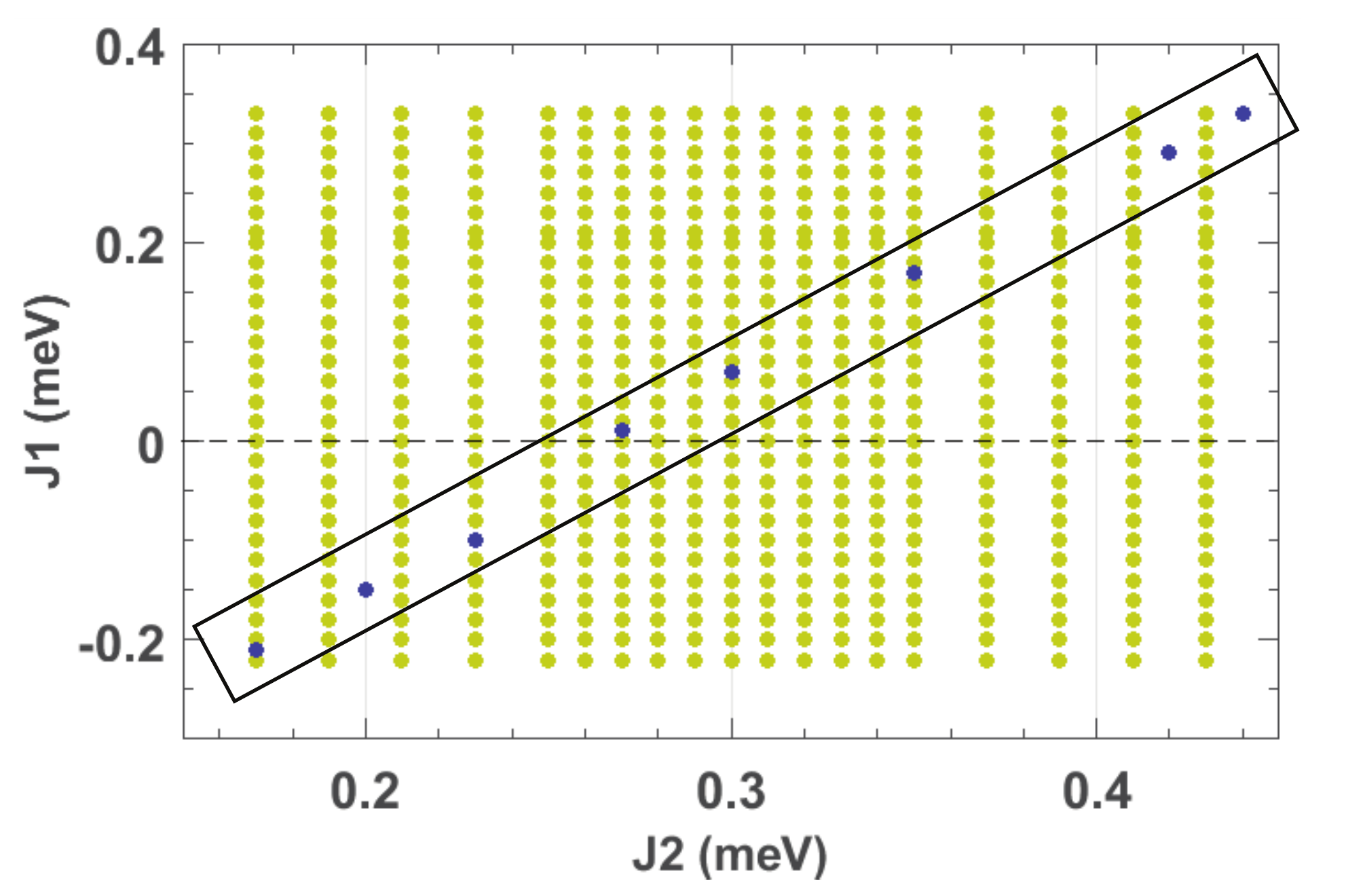}
	\caption{\label{PhaseDiagram}Range of possible $J_1$ and $J_2$ interaction strengths in \ce{BaCuTe2O6} for $J_3=2.9$~meV and $J_6=0.035$~meV, obtained by comparing the INS data to spin-wave simulations. 
	%The yellow points represent the values of the interactions that were tested. 
	A large number of combinations of $J_1$ and $J_2$ were generated and tested (yellow points). 
	%The blue points represent the $J_1$-$J_2$ combinations that reproduce the INS data. 
	Combinations lying within the black rectangular box showed good agreement with the INS spectrum, indicating the region of allowed $J_1$ and $J_2$ values.
	The blue points represent combinations that were investigated in particular detail.
	%From this, the range of allowed values of $J_1$ and $J_2$ were deduced to lie within the back box.
	%The error bars indicate the range of $J_1$ values for the given $J_2$, $J_3$, $J_4$ and $J_6$ values from which an area in the $J_1$-$J_2$ space is found (blue box) that agrees well with the INS data.
	}
\end{figure}

The spin-wave simulations for \ce{BaCuTe2O6} were calculated using the spinW program \cite{Toth2015} with the intrachain coupling fixed at $J_3'=4.56$~meV. The interchain interactions $J_1$, $J_2$ and $J_6$ were included and all the interactions were assumed to be isotropic or Heisenberg which is usually a good approximation for Cu$^{2+}$ ions. The neutron scattering structure factor was simulated for a very wide range of combinations of both ferromagetic and antiferromagnetic values of the intrachain couplings. In all cases a complex multi-branch spin-wave spectrum was found. Many of the combinations could be instantly rejected because they were incompatible with the known magnetic structure of \ce{BaCuTe2O6} \cite{Samartzis2021}. The simulations from the remaining combinations were compared to various low energy cuts and slices at 2~K collected on the LET and FLEXX spectrometers such as those shown in Figures~\ref{fig:LETscEQCollection},~\ref{fig:LETscQQCollection} and \ref{fig:FLEXXscEQCollection}. 
A detailed visual comparison was performed of the data and simulation taking into account both the disperion energies and intensities to judge if there was a match.
The ranges of the exchange interactions found to be compatible with the data are described in Table~\ref{tab:INSInteractions} and lie within the rectangular area in Fig.~\ref{PhaseDiagram}.

We found that it is essential to have a finite and antiferromagnetic $J_2$ to explain the data, this is to be expected because of the 120$^{\circ}$ magnetic order around each corner-sharing triangle formed by this interaction. At the same time we found the values of $J_1$ and $J_2$ to be highly coupled.  $J_2$ can take values in the range $0.17<J_2<0.44$~meV. For a given $J_2$, $J_1$ was found to lie within the range $ 2\times J_2-0.56 < J_1 < 2\times J_2-0.52 $~meV. For the smallest value of $J_2=0.17$~meV, $J_1$ is ferromagnetic with size -0.20~meV, while for the largest value of $J_2=0.44$~meV, $J_1$ is antiferromagnetic of size 0.34~meV. The factor of two in the relation between $J_2$ and $J_1$ can be understood to arise from the fact that there are twice as many $J_2$ bonds as $J_1$ bonds so a two times larger change in $J_1$ is needed to influence the spectrum. Finally $J_6$ was found to have very little effect on the excitations and could take either ferromagnetic values or antiferromagnetic values with no observable change in the simulations. The insensitivity of the magnetic spectrum to $J_6$ can be understood as due to the fact that $J_6$ couples together parallel $J_3$ chains (see Fig.~\Ref{fig:interactions}(c)). Since there is a shift between neighboring chains, a Cu$^{2+}$ ion on one chain is coupled by $J_6$ to {\em two} Cu$^{2+}$ ions on each neighboring chain, and since these two Cu$^{2+}$ ions are antiparallel due to the strong antiferromagnetic $J_3$ interaction this coupling is frustrated for both ferromagnetic and antiferromagnetic values of $J_6$. As a consequence, the energy of the $J_6$ bonds effectively cancels. We note that the spins on neighboring chains coupled by $J_6$ are always perpendicular to each other in response to this frustration. It was not possible to identify a unique solution for the Hamiltonian of \ce{BaCuTe2O6} since the simulated spectrum changes only very gradually within the range of compatible solutions. Example combinations of exchange interactions that fit the data are listed in Table~\ref{tab:INSInteractions}.

\begin{table}
	\footnotesize
	\caption{\label{tab:INSInteractions}Allowed ranges of the exchange interaction of \ce{BaCuTe2O6} obtained by comparison of the INS spectrum to spin-wave theory. The units are meV.}
	\begin{center}
		\resizebox{\columnwidth}{!}{%
			\begin{tabular}{ |c|c|c|c|c|c| }
				\cline{1-6}
			    %				&&&&&\\
			     &   & \multicolumn{4}{c|}{ } \\
			    \multicolumn{1}{|c|}{$J$} & \multicolumn{1}{c|}{Allowed Values} & \multicolumn{4}{c|}{Examples} \\			    
			     &   & \multicolumn{4}{c|}{ } \\
				%				&&&&&\\
				\hline		
				\hline		
				%				&&&&&\\
				$J_3$&$J_3=2.90$ & $2.90$  & $2.90$ & $2.90$ &$2.90$\\
				 &$J_3'=\pi J_3/2 = 4.56$ &   &  &  &\\
				\hline
				%				&&&&&\\
				$J_2$&$0.17<J_2<0.44$&$0.17$&$0.27$&$0.30$&$0.44$ \\
				&&&&&\\
				\hline
				%				&&&&&\\
				$J_1$&$J_1<2J_2-0.52$&$-0.20$&$0.00$&$0.06$&$0.34$ \\				       
				& \& $J_1>2J_2-0.56$& & & &  \\
				%				&&&&&\\
				\hline
				%				&&&&&\\
				$J_6$& & & &$J_6>-0.12$&  \\
				& & & & \& $J_6<0.18$&  \\
				%				&&&&&\\
				\cline{1-6}
			\end{tabular}
		}
	\end{center}
\end{table}

The spin-wave spectra simulated for the solution $J_3'=\pi J_3/2$ where $J_3=2.90$~meV, $J_2=0.30$~meV, $J_1=0.05$~meV and $J_6=0.035$~meV are shown alongside the data in the lower panels of Figures~\ref{fig:LETPowderEQ},~\ref{fig:LETscEQCollection},~\ref{fig:LETscQQCollection} and \ref{fig:FLEXXscEQCollection}. To make a direct comparison to the data, the simulated structure factor was convolved with a Gaussian to mimic the effects of the energy resolution and the magnetic form factor of the Cu$^{2+}$ ion was included in the calculations. The same energy and wavevector integration ranges used for the LET data were also used in the simulations.
This was achieved by, averaging several simulated slices with different out-of-plane ([$K$,$-K$,0]) wavevectors within the out-of-plane integration range of the data, and where appropriate a similar averaging was performed over wavevector within the in-plane integration range ([$H$,$H$,0] or [0,0,$L$]) and over energy within the energy integration range. For the simulations of the FLEXX data, the wavevector resolution broadening was taken into account by taking the Gaussian weighted average over the expected resolution widths in the directions out-of-plane ([$K$,$-K$,$0$]) and in-plane perpendicular to each slice by $0.1$~\AA$^{-1}$ and $0.02$~\AA$^{-1}$ respectively.

In general, good agreement between the energies and intensities of the low energy magnetic excitations measured at $T=2$~K and the spin-wave calculations was accomplished. The experimental data appear somewhat broader that the simulations, we attribute this to partial deconfinement of the spin-waves into spinons with increasing energy.\\

\begin{figure*}
\includegraphics[width=\textwidth]{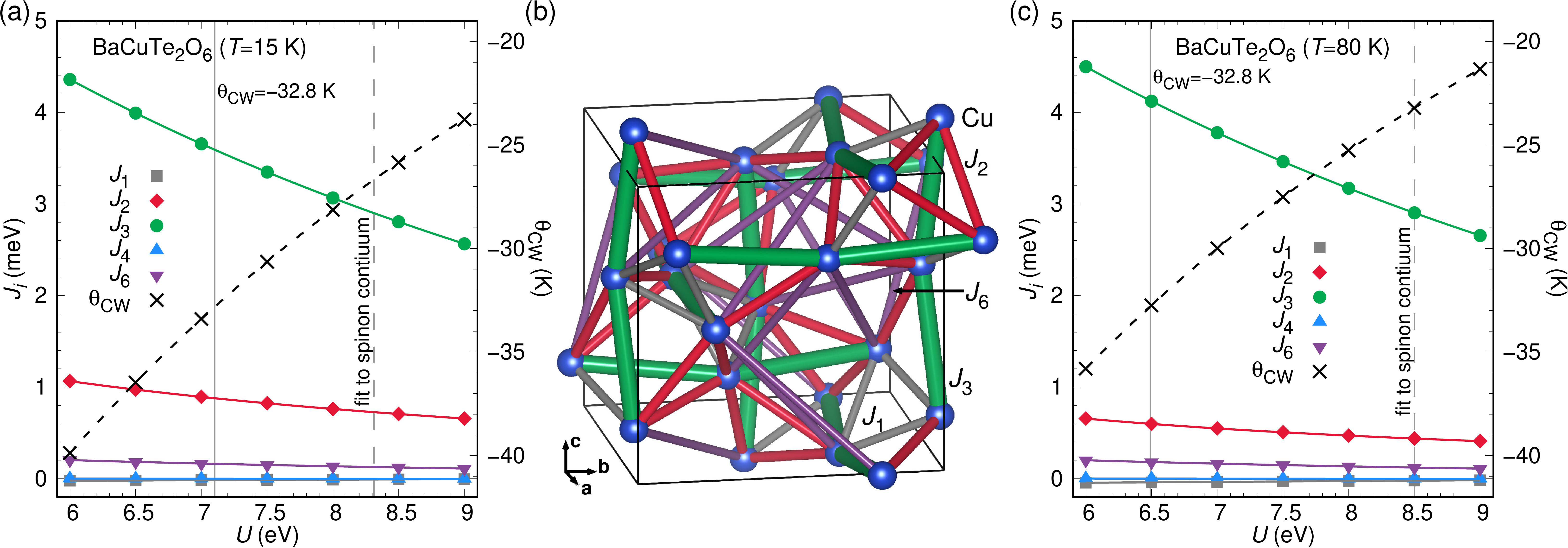}
\caption{\label{fig:exchange}(a) Exchange couplings of \ce{BaCuTe2O6} determined by DFT energy mapping for the structure determined at $T=15$\,K~\cite{Samartzis2021}. The solid vertical line marks the set of couplings picked out by the experimental Curie Weiss temperature, 
$T_{\rm CW}=-32.8$\,K giving $U=7.1$\,eV~\cite{Samartzis2021}. 
An alternative set of interactions was obtained by using the value $J_3=2.90$\,meV corresponding to $U=8.3$\,eV (dashed vertical line) which was extracted from the INS data. 
For a more detailed discussion of the choice of $U$ see the text. 
(b) Network formed by the exchange interactions $J_1$ (grey), $J_2$ (red), $J_3$ (green) and $J_6$ (purple). 
(c) Same as (a) but calculated for the \ce{BaCuTe2O6} structure determined at $T=80$\,K~\cite{Bag2021}.
}
\end{figure*}

\section{Density functional theory}

\subsection{Methods}

Density functional theory (DFT) provides an alternative way to determine the magnetic exchange interactions of a material based on its crystal structure. 
We performed all electron density functional theory calculations using the full potential local orbital (FPLO) code~\cite{Koepernik1999} employing the generalized gradient approximation (GGA) exchange and the correlation functional~\cite{Perdew1996}. For the electronic structure calculations, we used the $T=15$\,K crystal structure of \ce{BaCuTe2O6} which was determined in Ref.\ \cite{Samartzis2021} by neutron diffraction. For comparison we also used the $T=80$\,K crystal structure determined from X-ray diffraction \cite{Bag2021}. The symmetry was lowered to $P2_1$ in order to have six symmetry inequivalent Cu$^{2+}$ sites, allowing for 13 spin configurations with distinct energies, thus enabling the extraction of the exchange couplings $J_1$, $J_2$, $J_3$, $J_4$ and $J_6$ by energy mapping~\cite{Guterding2016,Iqbal2017}. 

\subsection{Results}
Figure~\ref{fig:exchange}\,(a) shows the strengths of the exchange interactions $J_1$, $J_2$, $J_3$, $J_4$ and $J_6$ resulting from the DFT energy mapping for the structure determined at $T=15$\,K using powder neutron diffraction~\cite{Samartzis2021}. Figure~\ref{fig:exchange}\,(c) shows the same couplings, but calculated for the x-ray diffraction structure determined at  $T=80$\,K~\cite{Bag2021}. 
%($J_4$ was found to be negligibly small).
%and the resulting geometry of the exchange network. 
Since the onsite Coulomb repulsion $U$ of the Cu $3d$ orbitals is unknown, the couplings were extracted for seven values of $U$ within the expected range of 6 to 9\,eV, while the Hund's rule coupling strength $J_{\rm H}$ was kept constant at 1\,eV as this has been shown to describe many Cu$^{2+}$ compounds correctly~\cite{Jeschke2013,Jeschke2015}. 
As there is no \emph{a priori} way to know the $U$ parameter, the energy mapping method needs additional information to fix the overall energy scale of the interactions. 
Usually, the experimental Curie Weiss temperature is used for this purpose. We can calculate it from the exchange couplings via
\begin{equation}
    T_{\rm CW}=-\frac{2}{3}S(S+1)\big(J_1+2J_2+J_3+J_4+2J_6\big)
\end{equation}
where $S=1/2$.
The value of $T^{\chi_{DC}}_{\rm CW}=-32.8$\,K extracted from DC magnetic susceptibility measurements ($\chi_{DC}$) for \ce{BaCuTe2O6} \cite{Samartzis2021} leads to the on-site potential $U=7.1$~eV for the $T=15$\,K crystal structure, indicated in Fig.~\ref{fig:exchange}\,(a) by the solid vertical gray line and corresponds to the set of couplings $J_1= -0.02(1)$\,meV, $J_2= 0.88(1)$\,meV, $J_3= 3.59(1)$\,meV, $J_4=0.00(1)$\,meV, $J_6= 0.16(1)$\,meV.  
The same criterion for the $T=80$\,K structure (Fig.~\ref{fig:exchange}\,(c)) corresponds to $U=6.5$~eV, yielding $J_1= -0.04(1)$\,meV, $J_2= 0.60(1)$\,meV, $J_3= 4.13(1)$\,meV, $J_4=0.00(1)$\,meV, $J_6= 0.18(1)$\,meV.  

In the case of \ce{BaCuTe2O6}, the comparison of 
%spin wave theory and 
the expression for the spinon continuum to the high energy INS spectrum as detailed in the previous section and Ref.~\cite{Samartzis2021}, provides another criterion for the overall energy scale by fixing the largest exchange interaction at $J_3=2.90$ meV. 
Evaluating the DFT calculation in this way leads to the Coulomb interaction $U=8.3$\,eV for the $T=15$\,K crystal structure (dashed vertical gray line in Fig.~\ref{fig:exchange}\,(a)), resulting in the set of exchange parameters $J_1=-0.01(1)$\,meV, $J_2=0.73(1)$\,meV, $J_3=2.90(1)$\,meV, $J_4=0.00(1)$\,meV, $J_6=0.13(1)$\,meV. 
The energy scale of these interactions is somewhat smaller that those based on $\chi_{DC}$ and yields $T^{INS}_{\rm CW}=-26.7$\,K. 
The same evaluation for the $T=80$\,K structure (Fig.~\ref{fig:exchange}\,(c)) yields $U=8.5$\,eV and $J_1=-0.02(1)$\,meV, $J_2=0.44(1)$\,meV, $J_3=2.90(1)$\,meV, $J_4=0.00(1)$\,meV, $J_6=0.12(1)$\,meV and corresponds to a Curie-Weiss temperature of $T^{INS}_{\rm CW}=-23.2$\,K. 
This last set of couplings is in excellent agreement with the calculation reported in Ref.~\onlinecite{Bag2021} for the same 80~K crystal structure. All these DFT results are listed for comparison in Table~\ref{tab:DFTInteractions}. One unusual feature of the Heisenberg Hamiltonian thus obtained is the fact that the shortest exchange path ($d_{Cu-Cu}=4.75$\,{\AA}) is found to have only a negligible exchange interaction $J_1$ which is less than 1\% of the dominant exchange $J_3$. This must be due to the involved geometry of the material; the shortest superexchange paths for both couplings are Cu-O-Ba-O-Cu, but the path for $J_1$ is much more bent than the path for $J_3$. However, in long paths like this, complex cancellation effects which are not accessible to simple arguments are possible. 

For a given structure, the two sets of exchange interactions are very consistent despite the difference in overall energy scale. The details of the energy dispersions will be governed by the ratios $J_2/J_3$ and $J_6/J_3$. For the $T=15$\,K structure, the ratios are almost the same, being $J_2/J_3=24.4\%$ and $J_6/J_3=4.5\%$ for the interactions based on the Curie-Weiss temperature from $\chi_{DC}$, and $J_2/J_3=25.0\%$ and $J_6/J_3=4.4\%$ for the interactions based on the INS $J_3$ value. The $T=80$\,K structure gives the ratios $J_2/J_3=14.5\%$ and $J_6/J_3=4.4\%$ for the $\theta_{\rm CW}$ criterion and $J_2/J_3=15.1\%$ and $J_6/J_3=4.2\%$ for the INS criterion.

% table in appendix?
% # U  J_1(K)  J_2(K)   J_3(K)   J_4(K)  J_6(K) TCW(K)            
% 6   -0.3(1) 12.4(1) 50.6(1) 0.0(1) 2.4(1) -39.9 # BaCuTe2O6, U=6eV, 4x4x4
% 6.5 -0.2(1) 11.3(1) 46.3(1) 0.0(1) 2.1(1) -36.5 # BaCuTe2O6, U=6.5eV, 4x4x4
% 7   -0.2(1) 10.4(1) 42.4(1) 0.0(1) 1.9(1) -33.4 # BaCuTe2O6, U=7eV, 4x4x4
% 7.5 -0.2(1) 9.6(1) 38.9(1) 0.0(1) 1.7(1)  -30.6 # BaCuTe2O6, U=7.5eV, 4x4x4
% 8   -0.1(1) 8.8(1) 35.6(1) 0.0(1) 1.6(1)  -28.1 # BaCuTe2O6, U=8eV, 4x4x4
% 8.5 -0.1(1) 8.2(1) 32.6(1) 0.0(1) 1.4(1)  -25.8 # BaCuTe2O6, U=8.5eV, 4x4x4

\section{\label{sec:discussion}Discussion}
The Hamiltonians derived from the DFT calculations for the two different crystal structures and the INS data are in good qualitative agreement. They both show \ce{BaCuTe2O6} to be dominated by an antiferromagnetic $J_3$ which results in quantum spin chains running parallel to the crystallographic {\bf a}, {\bf b} and {\bf c}-axes. The important subleading exchange is the hyperkagome interaction $J_2$ which couples the chains together and gives rise to long-range antiferromagnetic order. Other weaker interactions that may be present are $J_1$ and/or $J_6$. 

Table~\ref{tab:DFTInteractions} compares the DFT and INS results. Both the DFT interactions based on the Curie-Weiss temperature and on the INS $J_3$ value are listed. The accuracy of the value of Curie-Weiss temperature determined from the DC susceptibility data is of course important to determine the correct interactions by the first method. A significantly smaller value of $T^{\chi_{DC}}_{\rm CW}=-18.9$\,K was indeed previously found \cite{Bag2021} showing the possible range of this quantity. However, according to Fig.~\ref{fig:exchange}\,(a) and (c) this small $T^{\chi_{DC}}_{\rm CW}$ would imply, at $U> 9.5$~eV, values of onsite interaction $U$ that are unusually large for Cu$^{2+}$. On the other hand, an accurate value of $J_3$ is required to determine the interactions by the second method. $J_3$ was found by comparing the lower boundary of the spinon continuum observed in the high energy INS data to the des Cloiseaux-Pearson expression $(\pi\times J_3/2)\left|\sin(\pi H)\right|$ \cite{dCP}. It should be noted that this expression is for an ideal chain without interchain coupling. The presence of weak interchain coupling in \ce{BaCuTe2O6} would probably reduce the renormalisation factor of $(\pi/2)$ leading to an underestimation of $J_3$ although this effect should be small. 

The choice of criterion, $\theta_{CW}$ or $J_3$ influences the overall scale of the interactions while maintaining the ratios of these interactions. On the other hand, the two different structures considered have somewhat different ratios of the interactions, with $J_2 / J_3$ being larger in the case of the  $T=15$\,K structure than the $T=80$\,K structure. As the exchange paths for $J_2$ and $J_3$ are complicated and the two crystal structures are very similar, no simple explanation based on perturbation theory arguments can be found. 
It is clear that the DFT result are highly sensitive to the structure and this underlines the importance of precise experimental determination of oxygen positions in correlated oxides. It should be noted however that the present DFT result for the $T=80$\,K structure are very similar to those given in Ref.~\cite{Bag2021} which also used this structure.

The fitting of the low energy INS spectrum below $T_{\rm N}$ to spin-wave theory gives a range of possible solutions for the interchain interactions which agree equally well with the data, where the values of $J_1$ and $J_2$ were found to be highly coupled. For $J_3=2.9$~meV the constraints were $0.17<J_2<0.44$~meV and $ 2\times J_2-0.56 < J_1 < 2\times J_2-0.52 $~meV, while a weak value of $J_6$ (either AFM or FM) was possible but difficult to extract from the data. Two example solutions are given in Table~\ref{tab:DFTInteractions}. 
%one with $J_2$ at its maximum value of 0.44~meV and the other with $J_1 = 0.0$~meV. 
It should be noted that the excitations of frustrated magnets tend to be renormalized downward compared to spin-wave theory due to quantum fluctuations. Since the $J_1$ and $J_2$ interactions of \ce{BaCuTe2O6} are frustrated, when fitting the low energy and low temperature excitations to spin-wave theory, an underestimation of these interaction values could occur. 
%Thus the actual values of $J_1$, $J_2$ could be a bit larger than those presented in Table~\ref{tab:DFTInteractions}.

\begin{table}
	\footnotesize
	\caption{\label{tab:DFTInteractions} Our DFT results based on the Curie-Weiss temperature and the $J_3$ value from INS for the $T=15$\,K~\cite{Samartzis2021} and $T=80$\,K~\cite{Bag2021} crystal structures are compared with previous DFT results \cite{Bag2021}. Two possible sets of interactions extracted from the INS are shown for comparison, one with $J_2$ maximum and the other with $J_1=0.0$. All results are given in units of meV}
	\begin{center}
		\resizebox{\columnwidth}{!}{%
			\begin{tabular}{ |c|c|c|c|c|c| }
				\cline{1-6}
				\cline{1-6}
				\hline	
				& & & & &\\		    
				& $J_1$ & $J_2$ & $J_3$ & $J_4$ & $J_6$ \\			    
				% & & & & & &\\	
				\hline		
				%DFT & -0.1(1)  & 9.5(1)  & 38.8(1) & 0.0(1)  & 1.7(1) & K \\
				%$T^{\chi_{\rm DC}}_{\rm CW}$=-32.8K    & -0.01(1) & 0.82(1) & 3.34(1) & 0.00(1) & 0.15(1) & meV \\
				%\hline
				%DFT & -0.1(1)  & 8.4(1)  & 33.7(1) & 0.0(1)  & 1.5(1) & K \\
				%$J_3^{\rm INS}$=33.7K   & -0.01(1) & 0.72(1) & 2.90(1) & 0.00(1) & 0.13(1)  & meV\\
				\hline
				DFT (15\,K struct.)&   &  &  &  &  \\
				%DFT (15\,K struct.)&  -0.2(1) & 10.2(1) & 41.7(1) & 0.0(1) & 1.9(1) &K\\
				$T^{\chi_{\rm DC}}_{\rm CW}$=-32.8K & -0.02(1) & 0.88(1) & 3.59(1) & 0.00(1) & 0.16(1) \\ \hline
				
				DFT (15\,K struct.)&   &  &  &  &  \\
				%DFT (15\,K struct.)&  -0.1(1) & 8.4(1) & 33.7(1) & 0.0(1) & 1.5(1) &K\\
				$J_3^{\rm INS}$=33.7K & -0.01(1) & 0.73(1) & 2.90(1) & 0.00(1) & 0.13(1)\\ \hline
				
				DFT (80\,K struct.)&   &  &  &  &  \\
				%DFT (80\,K struct.)&  -0.5(1) & 7.0(1) & 47.9(1) & 0.0(1) & 2.1(1) &K\\
				$T^{\chi_{\rm DC}}_{\rm CW}$=-32.8K & -0.04(1) & 0.60(1) & 4.13(1) & 0.00(1) & 0.18(1) \\ \hline
				
				DFT (80\,K struct.)&   &  &  &  &  \\
				%DFT (80\,K struct.)&  -0.2(1) & 5.1(1) & 33.7(1) & -0.0(1) & 1.4(1) &K\\
				$J_3^{\rm INS}$=33.7K & -0.02(1) & 0.44(1) & 2.90(1) & 0.00(1) & 0.12(1) \\ \hline
				
				%DFT (80\,K struct.)& 0  & 5  & 34 & 0  & 2  \\
				DFT (80\,K struct.)&   &   &  &   &    \\
				Ref. \cite{Bag2021}  & 0 & 0.43 & 2.93 & 0 & 0.17 \\ 			\hline
				
				INS data &  &  &  &  &     \\
				%INS data & 3.9 & 5.1 & 33.7 & 0.0 & & K   \\
				e.g. 1: $J_2=0.44$ max & 0.34 & 0.44 & 2.90 & 0.0 &  \\ 				\hline
				
				INS data &  &  &  &  &      \\
				%INS data & 0.0 & 3.1 & 33.7 & 0.0 &  & K   \\
				e.g. 2: $J_1=0$ & 0.0 & 0.27 & 2.90 & 0.0 &  \\ 				\hline
				
				\cline{1-6}
			\end{tabular}
		}
	\end{center}
\end{table}

On a quantitative level, the INS results give the ratio of $J_2/J_3$ in the range 5.8 - 15.2~\%, which is lower than the DFT value of 25~\% for the $T=15$\,K structure, however it does include the 80~K structure DFT prediction of 15\%. Concerning the other interactions, DFT predicts $J_1$ to be close to zero while $J_6$ is weakly antiferromagnetic. In contrast INS finds a small but finite value of $J_1$ for most solutions while the value of $J_6$ is small and cannot be determined. There is one INS solution where $J_1$ is zero which requires $J_2/J_3 = 7.5$\% - a ratio somewhat lower than the DFT predictions which however can be explained by the presence of quantum fluctuations (see discussion above). On the other hand, taking the maximum allowed INS ratio of $J_2/J_3 = 15.2$\% results in an AFM $J_1$ of 0.34~meV somewhat larger than the DFT prediction (see Table~\ref{tab:DFTInteractions}).

\ce{BaCuTe2O6} can be compared to its isostructural sister compounds SrCuTe$_2$O$_6$ and \ce{PbCuTe2O6}. While for \ce{PbCuTe2O6} the dominant antiferromagnetic $J_1$ and $J_2$ interactions lead to a highly frustrated possible spin liquid ground state, in SrCuTe$_2$O$_6$ the chain interaction $J_3$ is AFM and dominant as we find here for \bacu . For SrCuTe$_2$O$_6$, $J_3 \approx 4.22$\,meV \cite{Ahmed2015,Chillal2020_2,Koteswararao2015,Chillal2021,Bag2021}  while the value is smaller at 2.90\,meV for \ce{BaCuTe2O6} \cite{Samartzis2021,Bag2021}. 
In SrCuTe$_2$O$_6$, the subleading interactions that couple the chains together and bring long-range magnetic order in the $1\times\Gamma_1^1$ irreducible representation, are either the combination of an AFM $J_1$ with a weaker FM $J_2$ \cite{Chillal2021} or a FM $J_2$ and an AFM $J_6$ \cite{Chillal2021,Bag2021}. In contrast, for \ce{BaCuTe2O6} the long-range order forms in the $2\times\Gamma_2^1$ irreducible representation \cite{Samartzis2021} which we show here to be stabilized by an AFM $J_2$ interaction while weaker $J_1$ or $J_6$ interactions may also be present. \\

\section{Summary and Conclusions}

In summary, we explored the magnetic excitations of the quantum spin chain antiferromagnet \ce{BaCuTe2O6} using high resolution INS and found that the spinon continuum at the high and medium energy ranges gives way, at low energies and temperatures, to a complex series of spin-wave modes. While the spinon continuum fixes the values of the $J_3$ intrachain exchange constant, a spin-wave fit taking a renormalized $J_3$ value into account, successfully models the low energy spectrum. Good overall agreement was achieved for many energy and wavevector directions, by taking into consideration the energy resolution and integration ranges. The $J_2$ hyperkagome exchange interaction was shown to be the sub-leading antiferromagnetic term responsible for coupling the chains together, and a range of allowed antiferromagnetic values was found. $J_1$ was shown to be weaker and highly coupled to $J_2$ while $J_6$ could not be determined. 

The exchange interactions were also calculated by DFT. In agreement with INS, DFT found that $J_3$ is antiferromagnetic and dominant and the next strongest interaction is an antiferromagnetic $J_2$, a weak $J_6$ was also predicted while $J_1$ was close to zero. These calculations were surprisingly sensitive to the crystal structure so that the small difference between the structure at 80~K determined from X-ray diffraction and the 15~K structure from neutron diffraction resulted in a 40\% change in the ratio of $J_2/J_3$. Altogether, a qualitative picture of \ce{BaCuTe2O6} emerges of a spin chain compound with frustrated antiferromagnetic hyperkagome interchain couplings, different from SrCuTe$_2$O$_6$ where these interactions are ferromagnetic or \ce{PbCuTe2O6} which is highly frustrated due to dominant antiferromagnetic $J_1$ and $J_2$.

In conclusion, the $A$CuTe$_2$O$_6$ family shows a surprising sensitivity to the $A$-site ion and crystal structure, and continues to surprise us with the varied range of Hamiltonians that it harbours. We hope a deeper understanding of the structure-magnetic property relationship will be achieved in the future.\\

\begin{acknowledgments}
We thank Jean-Sébastien Caux for his calculation of the dynamical structure factor of the spin-1/2 Heisenberg antiferromagnetic chain. B.L. acknowledges the support of Deutsche Forschungsgemeinschaft (DFG) through project B06 of SFB 1143: Correlated Magnetism: From Frustration To Topology (ID 247310070). The powder synthesis, crystal growth and physical properties measurements took place at the Core Lab Quantum Materials, Helmholtz Zentrum Berlin für Materialien und Energie, Germany. We gratefully acknowledge the Science and Technology Facilities Council (STFC) for access to neutron beamtime at the LET ISIS facility and also for the provision of sample preparation. 
\end{acknowledgments}

%\bibliography{bacute2o6}


\end{document}